# Cosmic ray hydrodynamics at shock fronts

P. Duffy[1], L.O'C. Drury[2], H. Völk[1]*

[1] Max Planck Institut für Kernphysik, 69117 Heidelberg, Federal Republic of Germany
[2] Dublin Institute for Advanced Studies, 5 Merrion Square, Dublin 2, Ireland

**Abstract.** When a magnetised, thermal plasma containing a shock wave efficiently accelerates Cosmic Rays (CR) a reaction is exerted on the fluid structure. The simplest description of this process involves three conservation equations for the background fluid, with the CR pressure gradient included as an extra force, plus a hydrodynamic equation for the CR energy density. Since the CR spectrum is not determined in this approach two quantities are introduced, the ratio of CR pressure to energy density and a mean diffusion coefficient, that must be specified to close the system of equations. Physical arguments are presented for how these quantities should evolve and then compared with time dependent numerical solutions of acceleration at a plane parallel, piston driven shock.

## 1. Introduction

There are a variety of astrophysical environments where the energy density of Cosmic Rays (CRs) is comparable to that of the local thermal plasma. Such is the case, for example, in the interstellar medium (ISM) where the dynamics of the plasma is coupled to that of the energetic particles not only through the latter's pressure and energy flux but also by collective wave-particle excitation. The combined dynamics of thermal and non-thermal particles is important not just for the propagation of CRs in the ISM but also during the CR acceleration process itself which is believed to occur, at least up to the knee of the CR spectrum, at the strong shock waves of supernova remnants (SNRs) by the first order Fermi mechanism (Axford et al. 1977, Krymsky 1977, Bell 1978 a,b and Blandford and Ostriker 1978). The energetic particles are scattered by the magnetic irregularities which are essentially frozen into the upstream and downstream flows and gain energy by compression at the shock front. Once the pressure of the accelerated CRs rises to a significant fraction of the gas pressure their effect on the hydrodynamics becomes important so that a pure thermal gas dynamic approach to the evolution of the SNR is no longer accurate. The simplest way of calculating the hydrodynamic modification is to include the CR pressure, $P_C$, as an extra component in the *total* pressure. Staying with the example of a SNR, with the region ahead of the expanding shock front occupied by a diffusive tail of CRs the upstream plasma is accelerated, by this CR pressure gradient, and possibly heated, through the dissipation of resonantly excited Alfvén waves. The expanding blast wave then sweeps up not the undisturbed ISM nor the progenitor star's stellar wind but a mixture of a modified thermal plasma and non-thermal CRs. Furthermore since the energetic particles are scattered by the magnetic turbulence there is a diffusive coupling between the modified upstream precursor and the high pressure, low density interior of the remnant. With efficient particle acceleration the explosion energy of a supernova is therefore converted in a non-linear fashion not just into kinetic and thermal gas energies but also into non-thermal CR energy (Drury et al. 1989, Dorfi 1990, Jones and Kang 1990, Markiewicz et al. 1990 and Berezhko et al. 1993). This model of self consistent SNR and CR evolution can provide, with reasonable choices of parameters, both sufficient power to sustain the observed CR intensity and enough hot gas to explain the X-ray luminosity of young SNRs.

Crucial to the above picture, however, is exactly how the CR energy density, $E_C$, and pressure are to be calculated. In general these bulk quantities depend on an energetic particle spectrum that is neither in equilibrium nor a simple power law, since the shock structure is modified. This suggests that one should solve self consistently for the phase space density distribution function $f(x, p, t)$ of the energetic particles and from this obtain the fluid moments $E_C$ and $P_C$ (Falle and Giddings 1987, Bell 1987, Duffy 1992 and Berezhko et al. 1993). An alternative, and numerically simpler, approach which is suggested intuitively by the notion of a bulk CR pressure and energy density is to extend the gas dynamic view to include the CR population without obtaining information about the spectrum (Drury and Völk 1981 and Axford et al. 1982). The almost isotropic distribution of CRs is treated as a fluid with a negligible mass density but an important, and in some cases dominant, energy density. In contrast to the thermal energy density which, in a comoving volume element with dissipative heating ignored, can only change as a result of adiabatic compression the added effect of scattering off magnetic turbulence is included for $E_C$. While this picture of a compressible, diffusing fluid which provides an extra component to the total pressure is straightforward qualitatively, agreement with a kinetic solution of the CR transport is not so simple. Turning to compression first, as the volume element is deformed the

* P. Duffy   Internet duffy@peter.mpi-hd.mpg.de
  L.O'C. Drury Internet ld@cp.dias.ie
  H. Völk    Internet vlk@vampi.mpi-hd.mpg.de

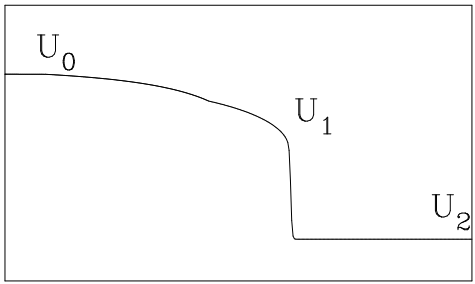

Gas velocity

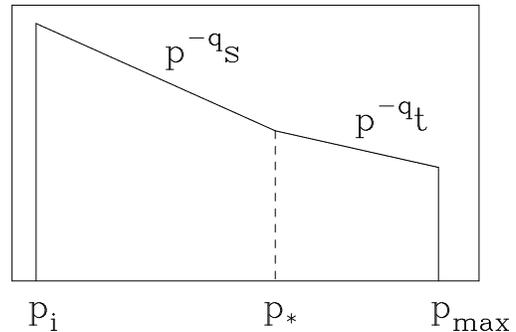

log(f(p)) vs. log(p)

**Fig.1a** Modified shock profile with $U_0$, $U_1$ and $U_2$ the far upstream, pre-shock and downstream flow speeds respectively.

**Fig.1b** The double power law spectrum with $q_s$ and $q_t$ given by the subshock and total compression ratios.

change in $E_C$ will depend on the CR pressure given by the equation of state, $P_C = (\gamma_C - 1)E_C$ where $\gamma_C$ is the CR 'adiabatic index'. During the acceleration process $\gamma_C$ will evolve from 5/3 to 4/3 as ultrarelativistic particles begin to dominate the spectrum. Secondly, since on a kinetic level the CR diffusion will be momentum dependent with a coefficient $\kappa(p)$, the rate of diffusion of the bulk energy density will be given by some mean $\bar{\kappa}$ of the microscopic coefficient weighted over the CR spectrum (see equation 3.2 below). Therefore CR hydrodynamics at shock fronts has the difficulty that the parameters $\gamma_C$ and $\bar{\kappa}$ depend on the evolving spectrum of energetic particles which is undetermined; the *closure problem*. Since, for the case of acceleration at a shock front, the solutions to the two fluid model are quite strongly dependent on the choice made for $\gamma_C$ and $\bar{\kappa}$ (Achterberg et al. 1984 and Heavens 1984) the challenge is to find physical arguments that would allow us to predict the evolution of the closure parameters in terms of fluid quantities.

The first attempt in this direction was contained in Drury et al. (1989) in connection with the simplified shell models. The expanding supernova remnant is divided into a number of regions with conservation laws describing the dynamics of each. Further work by Kang and Drury (1992) showed that these simplified and two fluid models agree reasonably well in their predictions. On the basis of sub-shock advection and injection Drury et al. estimate what proportion of the total CR energy density is made up of non-relativistic particles. Assuming that the remainder are ultra-relativistic this gives $\gamma_C$ as a weighted mean of 5/3 and 4/3. The mean diffusion coefficient is calculated by first obtaining the time dependent upper cut-off to the spectrum, $p_{max}$. A fraction of $\kappa(p_{max})$ is then taken for $\bar{\kappa}$. Two points can be made about these arguments. Firstly the closure parameters depend on the CR spectrum so that it would be more direct to relate $\gamma_C$ and $\bar{\kappa}$ to anything that could be said about modified CR distributions. Second is that any such arguments should be tested against a full kinetic code where available. It was not until Völk et al. (1991) (hereafter referred to as VDDM) that an attempt was made to construct a model with these points in mind. Use was made of the fact that in all but the completely cosmic ray dominated case, a sharp gas shock is modified to give a weaker subshock followed by a precursor in the upstream region caused by the CR pressure gradient (figure 1a). With a diffusion coefficient increasing with energy the spatial scale of this precursor is determined by the diffusion lengthscale of a particle with some momentum, $p_*$, intermediate between the point of injection, $p_i$, and the time dependent upper cut-off, $p_{max}$ (eq. 8 of VDDM). Consequently while those particles near injection only feel the compression due to the gas subshock, the highest energy CR see the entire subshock-precursor structure as a discontinuity. In accordance with test particle theory it was assumed that the spectral slopes at $p_i$ and $p_{max}$ are then determined by the subshock and total compression ratios respectively. With each of these slopes interpolated to the intermediate momentum $p_*$ this gives a double power law model for the modified CR distribution function from which the closure parameters may be calculated (figure 1b). Jones and Kang (1992) and Kang (1993) have used the time evolution of a test particle distribution function to calculate the closure parameters using equation (38) of Drury et al. (1989).

The double power law idea can be tested in the planar case through a direct comparison with a numerical kinetic solution to the problem. This is carried out below in section (2) where care is taken to consider a case for which the prediction of the closure parameters ought to be difficult. To this end we consider the limit of quasilinear theory, the Bohm limit, where the mean free path of a CR particle is equal to its gyroradius in the mean magnetic field. This ought to be the case near strong shocks where wave self-excitation may be so strong that diffusion approaches this lower bound of $\kappa \propto pv$, with $v$ the particle's velocity. Therefore $\bar{\kappa}$ will be a non-trivial average of a strongly momentum dependent quantity weighted by a spectrum extending over several decades of energy. Secondly we consider shock waves that are substantially modified, though not smoothed out completely, by the non-thermal population. Finally in a planar shock with losses ignored, acceleration will continue indefinitely so that $p_{max}$ enters the ultrarelativistic regime and $\gamma_C$ will come close to 4/3. We therefore restrict ourselves to intermediate timescales, i.e. timescales where most of the CR energy density is due to particles in the trans-relativistic regime so that $\gamma_C$ is not close to its lower or upper limit. It will be found (section 2) that for strongly mod-ified shocks in the Bohm limit during intermediate timescales that although the double power law approximation holds for the lowest energy CRs and although the spectrum flattens relative to the subshock compression at high energies the discrepancy between the model and the computed spectra is too great for the former to accurately predict the closure parameters. However, there is recent evidence (Berezhko et al. 1993 and E.Berezhko, private communication) that for long timescales, beyond the timescales considered here, the CR spectrum *does* relax to a double power law and that the discrepancies we find below are endemic to our relatively small dynamical range. Discussion of these results is left to the end of this paper after we have addressed what appears to be the most difficult part of the closure problem; accurately predicting $\gamma_C$ and $\overline{\kappa}$ for intermediate timescales (section 3). Comparison with the steady state Monte Carlo simulations due to Ellison and Eichler (1984) is less fruitful because of the intrinsically time dependent characteristics of the problem under consideration.

## 2. Kinetic description

If a plasma contains magnetic irregularities that effectively scatter high energy particles in the vicinity of a shock front then these CR can be accelerated to the point where they have an important influence on the shock structure itself (the reader is referred to the reviews by Drury 1983, Blandford and Eichler 1987, Berezhko and Krymsky 1988 and Jones and Ellison 1991 for the details of shock acceleration not discussed here). With self consistent wave generation ignored, i.e. the spectrum of fluctuations causing the scattering is assumed given, the thermal component is described by the three conservation equations for an ideal fluid, containing the added force exerted by the CRs. The transport equation for the isotropic part of the energetic particle phase space density, $f(x,p,t)$, describes the CR interactions on a microscopic level. In one spatial dimension these relations become

$$\frac{D\rho}{Dt} = -\rho \frac{\partial U}{\partial x} \quad (2.1)$$

$$\frac{DU}{Dt} = -\frac{1}{\rho}\frac{\partial}{\partial x}(P_G + P_C) \quad (2.2)$$

$$\frac{DE_G}{Dt} = -\gamma_G E_G \frac{\partial U}{\partial x} \quad (2.3)$$

$$\frac{Df}{Dt} = \frac{1}{3}\frac{\partial U}{\partial x} p \frac{\partial f}{\partial p} + \frac{\partial}{\partial x}\left(\kappa \frac{\partial f}{\partial x}\right) \quad (2.4)$$

where $\rho$, $U$, $E_G$ and $P_G$ denote the fluid's mass density, mass velocity, internal energy density and pressure with the time derivatives on the left hand side taken in a frame comoving with the fluid. With a non-relativistic equation of state for the gas, $P_G$ and $E_G$ are related by

$$P_G = (\gamma_G - 1)E_G$$

where $\gamma_G = 5/3$. Equation (2.4) describes how the CR distribution function changes as a result of adiabatic compression and scattering off the magnetic turbulence. The CR pressure and energy density are given by appropriate moments of the calculated spectrum

$$P_C(x,t) = \frac{4\pi}{3}\int p^3 v f(x,p,t)\, dp \quad (2.5)$$

$$E_C(x,t) = 4\pi \int p^2 T(p) f(x,p,t)\, dp \quad (2.6)$$

with $T(p)$ the kinetic energy of a particle of momentum $p$. The local time dependent CR adiabatic index is then determined by $\gamma_C(x,t) = 1 + P_C/E_C$.

This set of coupled partial differential equations can only be solved numerically. We adopt the method of Duffy (1992) which uses a finite difference scheme for equations (2.1)-(2.4). The numerical details are not repeated here. Physically we consider a piston driven shock with no initial CRs present. The process of injection whereby thermal particles are given an initial boost to suprathermal energies, where their gyroradius is larger than the shock thickness, is not yet understood and is therefore modelled in the following, simple manner. A fraction, $\eta$, of the upstream gas particles incident on the shock per unit time are injected as CRs. These fresh CRs have a single momentum $p_i$ which is fixed at a value of $0.1mc$ throughout this paper with $m$ and $c$ standing for the rest mass of an individual (single species) CR particle and the speed of light respectively. The injection profile is therefore a delta function in phase space; at $p_i$ and the instantaneous position of the shock. Once the CRs appear they diffuse back and forward across the shock thus gaining energy by compression. The momentum dependence of $\kappa$ determines the range of diffusion lengthscales ($L \sim \kappa/U$) and acceleration timescales ($\tau \sim \kappa/U^2$) from $p_i$ up to the highest energy particles. It would therefore be a good test of any model of the spectrum dependent closure parameters to choose a $\kappa(p)$ that varies strongly with momentum since in this case both $L$ and $\tau$ will differ by several orders of magnitude over the energy range of the CR spectrum. To this end we address that limit of CR transport theory (Skilling 1975) where strong wave excitation results in significant particle scattering after each gyration in the mean magnetic field. Thus the mean free path $\lambda$ is equal to the particle gyroradius; the Bohm limit. This gives $\kappa \propto pv$ so that the diffusion coefficient scales like $p^2$ ($p$) for non-relativistic (highly relativistic) particles.

Before presenting a typical calculation it is worth noting briefly how this system behaves in the, analytic, test particle limit; i.e. with the CR pressure gradient omitted from equation (2.2). With the initial conditions described above, a piston moving supersonically into an undisturbed and uniform medium, the gas dynamic solution is simply one of a shock wave that moves away from the piston separating uniform upstream and downstream states. A power law CR spectrum $f(p) = f_o(p/p_i)^{-q}$ is produced at the shock from injection up to some time dependent cut-off $p_{\max}$

$$\frac{dp_{\max}}{dt} = \frac{p_{\max}}{\tau_a} \quad (2.7)$$

with the acceleration timescale given by

$$\tau_a = \frac{3}{U_1 - U_2}\left(\frac{\kappa_1}{U_1} + \frac{\kappa_2}{U_2}\right)$$

where flow velocities in the shock rest frame are denoted by $U$ and the subscripts 1 and 2 refer to the upstream and downstream regions respectively. The spectral index is related to the shock compression ratio, $r = U_1/U_2$, by $q = 3r/(r-1)$ and the amplitude of the distribution function at $p_i$ is given by

$$f_o = \frac{r(q-3)\rho_1 \eta}{4\pi m p_i^3} \quad (2.8)$$

with $\rho_1$ the upstream mass density. While the downstream CR distribution is uniform, and equal to that at the shock, the upstream spectrum falls off exponentially with a lengthscale $\kappa/U$.

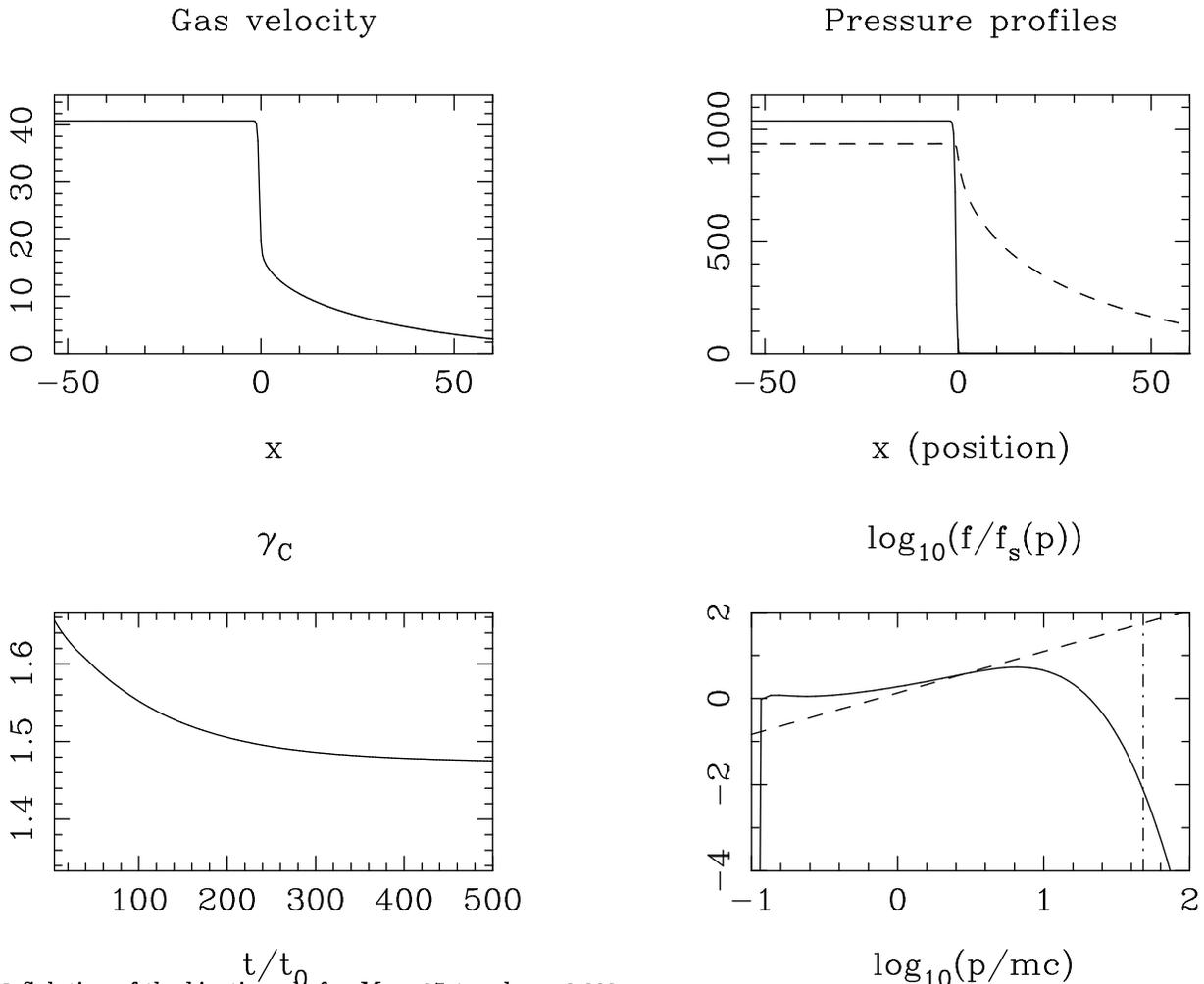

**Fig.2** Solution of the kinetic code for $M_p = 27.1$ and $\eta = 0.002$.

With the reaction included, the CR pressure gradient in the upstream region accelerates the plasma ahead of the shock to give a weaker subshock led by a broad precursor region. In figure 2 the first two panels show the velocity and pressure profiles for a piston Mach number of 27 and $\eta = 0.002$. These plots are taken after the bulk fluid properties have reached a quasi-steady state where, although the CR spectrum at the high energy end still continues to evolve very slowly, the energy densities do not change appreciably (Falle and Giddings 1987, Duffy 1992). The evolution of the first, numerically calculated, closure parameter is shown in the third panel, with $t_0$ the test particle acceleration timescale at injection. Initially $\gamma_C$ (at the shock) is close to 5/3 since the only particles present are those with momenta close to $p_i$. The CR adiabatic index decreases thereafter as the proportion of relativistic particles increases. Eventually, since the acceleration timescale goes at least linearly with momentum, it becomes harder to increase the upper cut-off and $\gamma_C$ flattens off.

The fourth plot shows the final CR spectrum at the shock. It is normalised throughout to $f_s(p) = f_0(p/p_i)^{-q_s}$ which is the test particle spectrum that would be expected from subshock compression alone; $f_0$ is given by equation 2.8 applied to the subshock and $q_s = 3r_s/(r_s - 1)$ with $r_s$ the subshock compression ratio. Consider first those particles near injection. They have relatively small values for $L$ and $\tau$. Consequently, as the initially strong shock evolves to give a weaker subshock plus a precursor, they are accelerated sufficiently quickly and over a small enough spatial scale to feel only the compression due to the evolving subshock. The spectrum near injection is therefore given, in amplitude and slope, by $f_s(p)$ which is evident from the normalisation used in the figure. Although perhaps not unexpected that test particle theory should apply to the lowest energy particles near the discontinuity it is by no means a trivial result and confirms the predictions of the double power law model of VDDM for the lowest energy particles. The spectrum is not so simple towards the higher end; here a single power law does not exist. The shock and precursor region evolve on the same timescale as that of the CR pressure which is determined by some intermediate momentum between $p_i$ and the upper cut-off. Since the highest energy particles are accelerated on a longer timescale than that of the fluid's evolution it follows that they cannot, unlike the CRs near $p_i$, fully adjust to the instantaneous fluid profile. It is therefore apparent that the shape of the spectrum at the highest energies is determined by the entire history of the subshock and precursor. These arguments can be tested by comparing the spectrum shape with that expected from the double power law argument $q_t = 3r_t/(r_t - 1)$ (dashed line). Although the spectrum flattens towards higher energies the slope is nowhere quite as flat as might be expected from the total subshock-precursor compression. In addition the rollover of the spectrum towards the highest energies is not described by a cutoff with a $q_t$ spectrum (dot-dashed line). While this was expected it further weakens the quantitative applicability of a double power law approximation so that for strong shocks with a strongly momentum dependent $\kappa$ the double power law model will not suffice as a tool for calculating the closure parameters in the fluid limit; the double power law spectrum will always overestimate the amount of highest energy CRs and will therefore give a lower value for $\gamma_C$. As an alternative it appears better to describe the overall spectrum by a power law at low momenta plus a bump at high momenta. We shall now construct a model for $\gamma_C$ and $\bar{\kappa}$ that is consistent with the generic features of such spectra.

## 3. Two fluid model and the closure problem

The CRs can be quantitatively described as a fluid by taking the energy density moment of equation (2.4) to give

$$\frac{DE_C}{Dt} = -\gamma_C E_C \frac{\partial U}{\partial x} + \frac{\partial}{\partial x}\left(\bar{\kappa}\frac{\partial E_C}{\partial x}\right) \quad (3.1)$$

The adiabatic compression term now depends on the CR equation of state through $\gamma_C$. The equations for $E_G$ and $E_C$, (2.3) and (3.1), differ through the diffusion term for the CRs. The mean coefficient $\bar{\kappa}$ is then related to $\kappa(p)$ and the gradient of the spectrum by

$$\bar{\kappa}(x)\int p^2 T(p)\frac{\partial f}{\partial x}dp = \int \kappa(x,p)p^2T(p)\frac{\partial f}{\partial x}dp \quad (3.2)$$

While we have gained, through the introduction of (3.1), a simpler description of the coupled system this has been done at the expense of introducing the two undetermined closure parameters. With the numerical solution of the previous section as motivation the problem is now to construct an argument for the evolution of $\gamma_C$ and $\bar{\kappa}$ in terms of only those variables that enter the two fluid model.

With a view to modelling integral moments of the CR distribution, as opposed to predicting the exact distribution itself, our scheme essentially consists of approximating the *power law and bump* shape of a self consistent spectrum as shown in figure 2 with a *power law and spike*. In other words the distribution function at the shock is written as

$$f(p,t) = f_0\left(\frac{p}{p_i}\right)^{-q_s} + f_1\delta(p-p_b) \quad (3.3)$$

The first term is the subshock power law distribution which is taken to extend from $p_i$ to the upper cut-off $p_{\max}(t)$ obtained by integrating equation (2.7) between the far upstream and downstream states over the fluid's history and based on the total compression ratio $q_t$. The effect of the enhanced distribution at high energies is contained in the delta function taken at $p_b$ which is some constant fraction of the upper cut-off, $\alpha = p_b(t)/p_{\max}(t)$, the one free parameter in this approach. While equation 3.3 may appear rather primitive it contains all the predictable elements of the low energy part and makes only the simplest assumptions about the high energy spectrum. As a model for the spectrum of the highest energy particles equation (3.3) is quite crude but we are more interested here in predicting the closure parameters which are integral quantities as far as the momentum dependence is concerned.

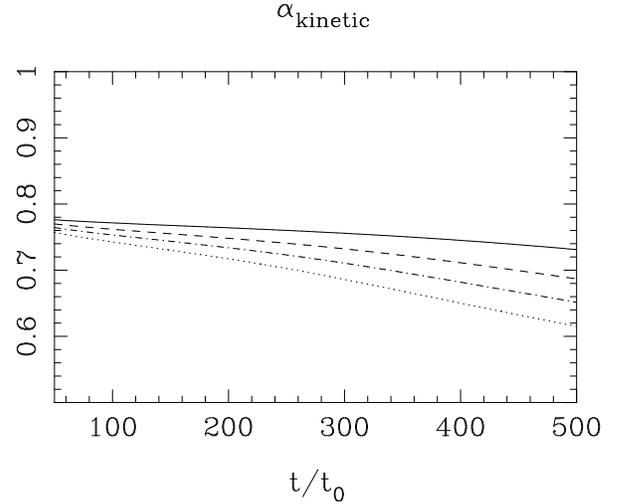

Fig.3 Evolution of $\alpha_{\text{kinetic}} = p'(t)/p_{\max}(t)$ for $M_p = 17.81$ (solid), 23.25 (dashed), 27.1(dot-dashed) and 31.97 (dotted).

We can immediately write down zeroth order approximations to the CR energy density and pressure as predicted from the test particle subshock spectrum

$$E_0 = 4\pi f_0 \int_{p_i}^{p_{\max}(t)} p^2 T(p)\left(\frac{p}{p_i}\right)^{-q_s} dp \quad (3.4)$$

$$P_0 = \frac{4\pi}{3}f_0 \int_{p_i}^{p_{\max}(t)} p^3 v \left(\frac{p}{p_i}\right)^{-q_s} dp \quad (3.5)$$

In a two fluid calculation either equation (3.1) or one like it is solved to give $E_C$. The difference, $E_1 = E_C - E_0$, is therefore the energy density of the extra component to the spectrum at high energies. By assumption in equation (3.3) this addition is entirely due to the delta function contribution near the upper cut-off so that the corresponding correction to the zeroth order pressure is

$$P_1 = (\gamma_1 - 1)E_1 \quad (3.6)$$

where $\gamma_1 = 1 + p_b v_b/3T(p_b)$ is the adiabatic index of a delta function distribution at $p_b$. An alternative but equivalent way of looking at this is that $E_1$ determines the scale height $f_1$ by

$$f_1 = \frac{E_1}{4\pi p_b^2 T(p_b)} \quad (3.7)$$

from which $P_1$ can also be calculated. Either way we may now write down a model adiabatic index at the shock

$$\gamma_m = 1 + \frac{P_0 + P_1}{E_0 + E_1} \quad (3.8)$$

that depends only on quantities which enter a two fluid model and containing only one free parameter $\alpha$ of order unity.

Turning to the mean diffusion coefficient it is not clear whether it is the most natural way to approximate $\bar{\kappa}$ at the shock and then use this value at all positions in order to solve the system of equations (2.1)-(2.3) and (3.1). Since $\bar{\kappa}$ does not connect local quantities (like $\gamma_C$ with regard to $P_C$ and $E_C$) but rather describes the diffusive character of the energetic particle distribution, a presumably more appropriate way is to use a spatial average for $\overline{\kappa}$ over the shock precursor in equation (3.1) to a first approximation, VDDM. To obtain such an average we spatially integrate eq. (3.2) from the subshock at $x = 0$ into the upstream medium over a distance $L$ where $f(x,p,t)$ and $E_C(x,t)$ go to zero

$$\int_0^L dx\, \overline{\kappa}\frac{\partial E_C}{\partial x} = 4\pi \int_0^L dx \int p^2\, dp\, T(p) \kappa \frac{\partial f}{\partial x} \qquad (3.9)$$

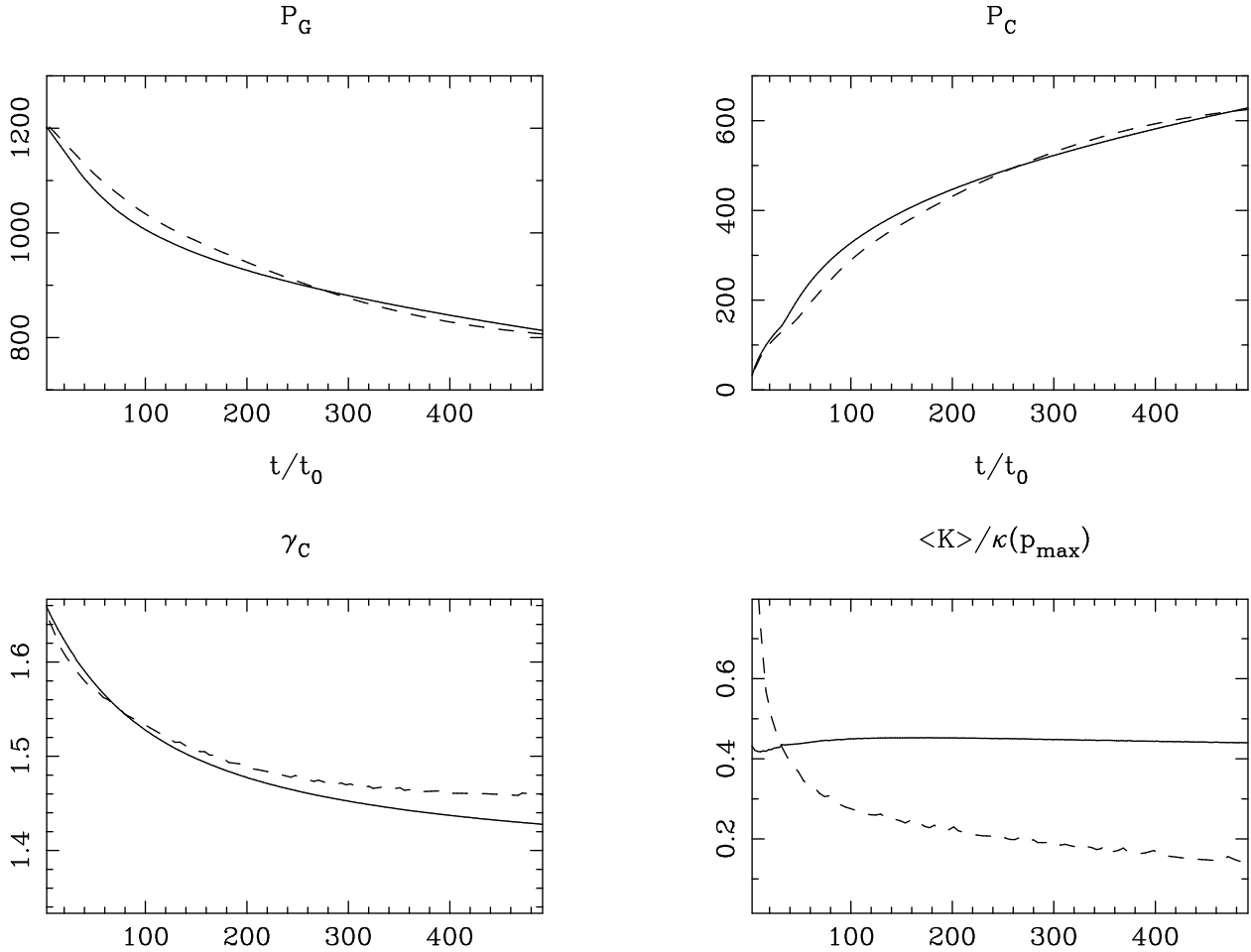

**Fig.4** Comparison of the fluid (solid) and kinetic (dashed) solutions for $M_p = 23.25$ and $\eta = 0.002$.

Defining the average $<\overline{\kappa}>$ by

$$\int_0^L dx\, \overline{\kappa}\frac{\partial E_C}{\partial x} = <\overline{\kappa}> \int_0^L dx\, \frac{\partial E_C}{\partial x} = <\overline{\kappa}> E_C(x=0,t)$$

and the precursor average $<\kappa>$ by

$$\int_0^L dx \int p^2\, dp\, T(p) \kappa \frac{\partial f}{\partial x} = \int p^2\, dp\, T(p) <\kappa> f(x=0,p,t).$$

Combining the above we obtain the required precursor averaged mean diffusion coefficient as the energy weighted mean of $<\kappa>$

$$<\overline{\kappa}> = \frac{4\pi \int p^2\, dp\, T(p) <\kappa> f_s(p,t)}{4\pi \int p^2\, dp\, T(p) f_s(p,t)} \qquad (3.10)$$

For the model spectrum of equation (3.3) and the diffusion coefficient written as $\kappa(p) = g(p)\kappa_0$, where $\kappa_0$ is the value at injection, the second closure parameter then becomes

$$<\overline{\kappa}>_m = \frac{1}{E_C}\left(g(p_b)E_1 + \int_{p_i}^{p_{max}} g(p)\, dE_0\right) \qquad (3.11)$$

where $dE_0 = 4\pi f_0 p^2 T(p)(p/p_i)^{-q_s} dp$ is the partial energy density of the zeroth order spectrum.

The coupled thermal and non-thermal system can now be solved numerically as a two fluid system using equations (2.1) to (2.3) for the gas coupled to equation (3.1) for the CRs with the closure parameters derived from the model spectrum (3.3) as described above. There is one remaining problem of what exact value ought to be taken for $\alpha$, the ratio of the $p_b(t)$ to $p_{max}(t)$, or indeed even if it is sensible to tie the evolution of the bump to that of the upper cut-off by taking $\alpha$ to be constant. Although this can only be settled by a direct comparison between the results of a two fluid model and kinetic solution (which is carried out below), the choice of $\alpha$ can be made less arbitrary by using the results of the full kinetic solution of the previous section. Here we can calculate the excess energy density in the modified distribution over the test particle subshock distribution, $\Delta E = E_C - E_0$, and similarly $\Delta P = P_C - P_0$. These differences are due to the flatter distribution towards the upper end of the spectrum in figure 2. We then ask at what momentum, $p'$, we would have to place a delta function distribution to obtain the correct ratio of $\Delta P/\Delta E$. The amplitude of the delta function can also be obtained but is of little practical interest to us since it is not needed for the closure parameters. We then have

$$\frac{p'v'}{T(p')} = 3\frac{\Delta P}{\Delta E}$$

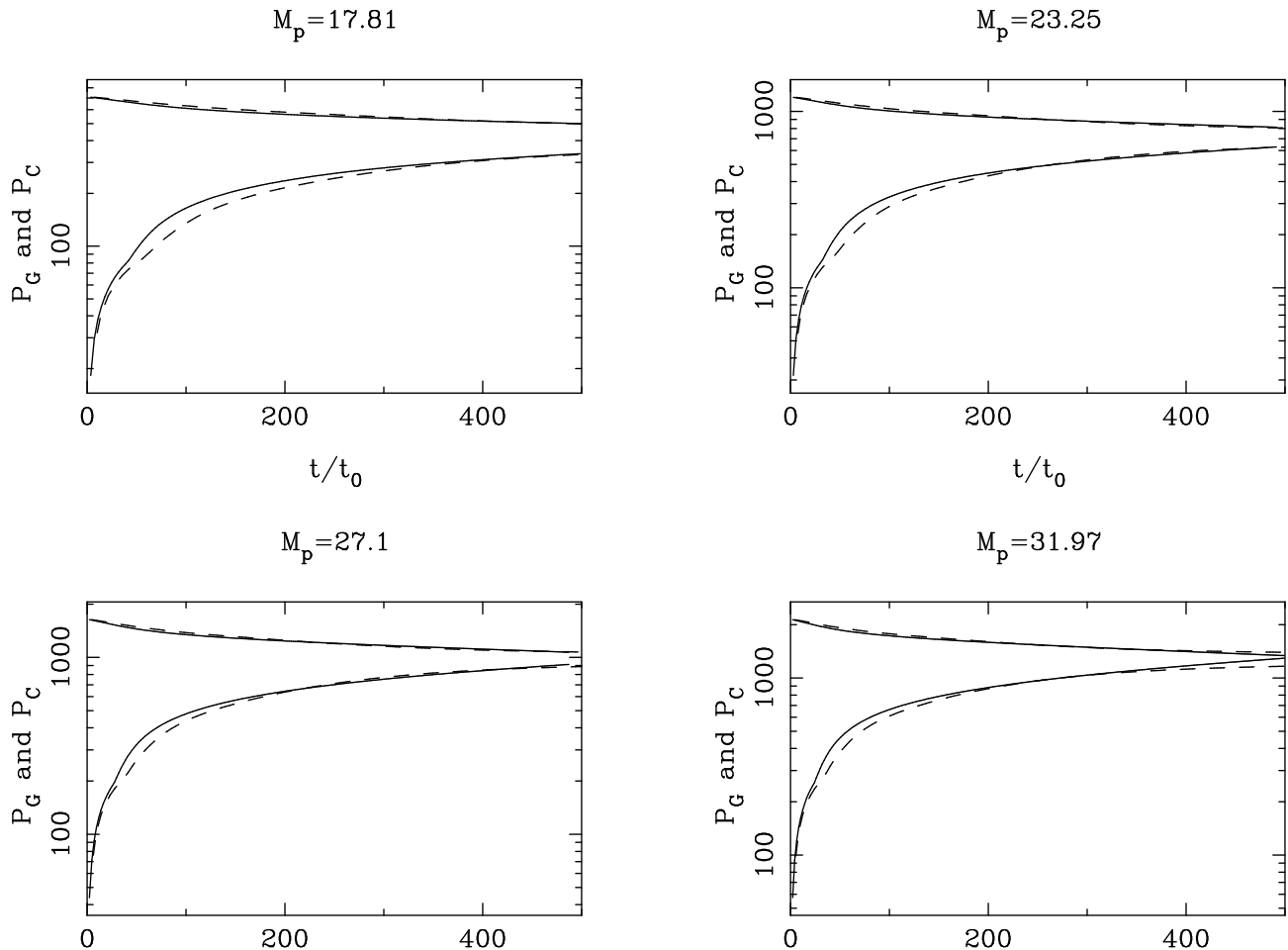

**Fig.5** Evolution of $P_G$ and $P_C$ for the fluid (solid) and kinetic (dashed) solutions with a range of Mach numbers and $\eta = 0.002$.

| $M_p$ | $\gamma_m/\gamma_C$ | $<\overline{\kappa}(t)>_m / <\kappa(x,t)>$ |
|---|---|---|
| 17.81 | 0.98 | 2.4 |
| 23.25 | 0.98 | 2.9 |
| 27.10 | 0.98 | 3.7 |
| 31.97 | 0.97 | 3.4 |

table 1

The evolution of $\alpha_{\text{kinetic}}(t) \equiv p'(t)/p_{\max}(t)$ is plotted in figure 3 for four different Mach numbers. The slow decline of $\alpha_{\text{kinetic}}$, which in part can be attributed to the fact that $p_{\max}(t)$ as calculated by (2.7) is an increasing overestimate for CR spectra at modified shocks, suggests a value of $\alpha = 0.6 - 0.8$ for the model spectrum (3.3).

The numerical solution of the two fluid system is compared with that of the kinetic code in figure 4 for a piston Mach number of 23.25 and $\eta = 0.002$ and for what is found to be the best value of $\alpha = 0.78$.

In each panel the solid and dashed lines show the evolution of quantities obtained from the fluid and kinetic codes respectively. The first two plots show the fall and rise of the thermal and non-thermal components of the total pressure at the gas subshock. The third panel compares the CR adiabatic exponent as given by equation (3.8) with that obtained from the CR spectrum *at the shock* with the two differing by no more than 2% throughout the evolution. The precursor averages of $\overline{\kappa}$ are shown in the fourth panel and differ by about a factor 3. What is most striking is that by replacing the local quantities $\gamma_C(x,t)$ and $\overline{\kappa}(x,t)$ with the spatially independent variables $\gamma_m(t)$ and $<\overline{\kappa}(t)>_m$ (eqns. 3.8 and 3.11), we have come close to reproducing the true, kinetic evolution of the bulk pressures with a two fluid model. This agreement is also found for a range of Mach numbers, at the same level of injection and with $\alpha = 0.78$, as shown in figure 5. Of the two closure parameters, the time asymptotic value of $\gamma_m$ always lies within 3% of the kinetic value at the shock which is a better agreement than that found for the precursor averaged diffusion coefficients which differ by a factor of between 2 and 4 (table 1). Given that we are comparing a two fluid model containing spatially constant closure parameters with a kinetic solution where $\gamma_C$ and $\kappa$ are local quantities, this level of disagreement is perhaps not too surprising. Figures 6a and 6b show how, for $M_p = 23.25$, small

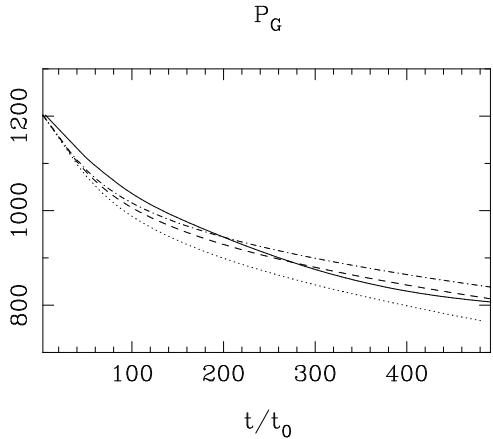

**Fig.6a** $P_G$ from the kinetic (solid) and fluid codes for $\alpha = 0.75$ (dot-dashed), 0.78 (dashed), 0.8 (dotted).

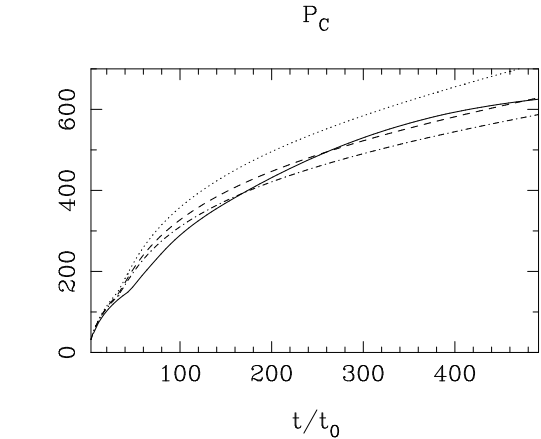

**Fig.6b** $P_C$ from the kinetic (solid) and fluid codes for $\alpha = 0.75$ (dot-dashed), 0.78 (dashed), 0.8 (dotted).

changes in the parameter $\alpha$ alter the evolution of the two fluid pressures.

## 4. Conclusions

We have used a model of CR spectra in time dependent modified shocks to predict values for the closure parameters of CR hydrodynamics at shock fronts. The lowest energy particles have a spectrum determined by the structure of, and injection at, the gas subshock. This accounts for almost all of the CR energy density at the subshock, the excess is due to more energetic particles that feel the extra compression in the precursor and have a flatter spectrum. We have looked at intermediate timescales where the most energetic CRs, whose acceleration timescale is so large that the slope of their spectrum cannot adjust to that given by the double power law model. For our dynamical, trans-relativistic, range (chosen as an intermediate case where $\gamma_C$ is not close to either of its limits) it is more accurate to attribute the excess energy to an individual particle momentum which evolves with the upper cut-off. The full hydrodynamic description of the coupled interaction between the thermal gas and the non-thermal CRs then agrees quite closely with that of a full kinetic treatment. As mentioned in the introduction there is evidence from a recent kinetic solution (Berezhko et al. 1993) that, for piston driven shock solutions, with an increased dynamical range the CR spectrum does eventually relax to the original double power law model.*
For timescales much longer than those considered here, where $\gamma_C$ eventually evolves towards 4/3 in the planar case, such a picture could then be used to calculate the closure parameters.

**Acknowledgements** We would like to thank J.Kirk and E.Berezhko for discussions.

## References

---

\* Ultimately the rate of particle acceleration at the shock front of a SNR will decay with decreasing Mach number to give low and high energy parts of the spectrum from particles accelerated during this final phase and older CRs that have been accelerated earlier.